\documentclass[a4paper]{VisionStyle}
\usepackage{epsfig}

\begin{document}

\title{XMM-Newton sets the record straight: No X-ray emission detected from PSR J0631+1036}

\author{J.\,Kennea\inst{1} \and F.\,C\'ordova\inst{1} \and
  S.\,Chatterjee\inst{2} \and J.\,Cordes\inst{2} \and C.\,Ho\inst{3} \and
  R.\,Much\inst{4} \and T.\,Oosterbroek\inst{4} \and A.\,Parmar\inst{4}}

\institute{
University of California, Santa Barbara, CA 93106, USA
\and
Cornell University, Ithaca, NY 14853-6801, USA
\and
Los Alamos National Labs, Los Alamos NM , 87545, USA
\and
ESTEC, Postbus 299, NL-2200 AG Noordwijk, The Netherlands
}

\maketitle 

\begin{abstract}
  
  The pulsar PSR J0631+1036 was discovered during a radio search of
  {\it Einstein} X-ray source error circles. A detection of a 288ms sinusoidal
  modulation in the {\it ASCA} lightcurve, the same period as the radio pulsar,
  appeared to confirm the association of the X-ray source and the pulsar.
  Its X-ray spectrum was said to be similar to that of middle aged
  $\gamma$-ray pulsars such as Geminga. However, an {\it XMM-Newton} observation
  of the PSR J0631+1035 field, along with a re-analysis of VLA data
  confirming the timing position of the pulsar, show a $75''$ discrepancy
  in location of the X-ray source and the pulsar, and therefore these
  cannot be the same object. The X-ray source appears to be the counterpart
  of an A0 star, detected by the {\it XMM-Newton} Optical Monitor. No 288ms
  period was detected from either the area around the pulsar or the bright
  X-ray source.  The upper limit on the X-ray luminosity with relation to
  the empirically observed correlation between radio measured dE/dt and
  X-ray luminosity is discussed.

\end{abstract}

\section{Introduction}

PSR J0631+1036 is a young ($\tau$=43,000yr), 288ms pulsar in the direction
of the Galactic anticentre. First reported by \cite*{jkennea-D1:z96}, it
was discovered in a radio search of the error circle of 2E 0628.7+1037, an
{\it Einstein} X-ray source.  PSR J0631+1036 has an unusually high
dispersion measure (DM) of 125 pc cm$^{-3}$, which corresponds to a
distance of 6.5 kpc (using the model of \cite{jkennea-D1:t93}). This
distance is the largest of the Galactic anticentre pulsars.  The
association of this source with the dark cloud LDN 1605
(\cite{jkennea-D1:l62}), however, suggests that this high DM is due to
ionised gas in or near the cloud. The actual distance to PSR J0631+1036 is
suggested to be $\sim1$kpc (\cite{jkennea-D1:z96}).

Although the radio pulsar lies 75" from the centre of the {\it Einstein} X-ray
error circle, which makes the position outside of the nominal 90\%
confidence area, \cite*{jkennea-D1:z96} argued that the error circle was
most likely underestimated due to it being partially occulted by an IPC
support strut. \cite*{jkennea-D1:z96} argue that the {\it ROSAT} source is also
partially occulted by the PSPC support strut, and use the same argument 
%as to the underestimating of the size of the X-ray error circle 
to explain the difference between the X-ray and radio positions. Ignoring
this argument, \cite*{jkennea-D1:bt97} report the source as not being
detected, and give an upper limit of $L_X = 1.7 \times 10^{33}$ erg/s
(0.1-2.4 keV).
%, ignoring the presence of 2E 0628.7+1037 in the {\it ROSAT} data.

It is clear from {\it ROSAT} and {\it Einstein} data that it is not
possible, using positional arguments alone, to determine whether PSR
J0631+1036 and 2E 0636.7+1037 are the same object, or whether the discovery
of the radio pulsar near the X-ray source was coincidental.

The apparently final piece of evidence needed to show that the X-ray source
and radio pulsar are related was given by the recently published work by
\cite*{jkennea-D1:t01}, in which analysis of an {\it ASCA} observation of
the PSR J0631+1036 field was presented.  \cite*{jkennea-D1:t01} report a
detection of the 288ms period in the {\it ASCA} data to within $10^{-7}$s
of the radio period.  In the paper they make no comment about the
discrepancy between the radio position and the X-ray position, but Torii
(2001, \textit{private communication}) does not consider a 75" discrepancy
to be particularly significant given the positional accuracy of {\it ASCA}.
Spectral fitting of the X-ray source lead \cite*{jkennea-D1:t01} to report
that PSR J0631+1036 has X-ray properties similar to $\gamma$-ray pulsars
such as Geminga.

In this paper we present analysis of an {\it XMM-Newton} observation of PSR
J0631+1036 utilising the EPIC-pn cameras.

\section{Observations}

The data presented are 14000s of {\it XMM-Newton} EPIC-pn data of which 8000 were
good (4000 lost due to high background), taken as part of the guaranteed
time program (PI: Fred Jansen).  No MOS data were obtained due to an
unusually high background level, and the RGS data were not useful due to
source brightness. The {\it XMM-Newton} Optical Monitor obtained 3000s of
simultaneous data with the UVW2 filter.

\begin{figure*}[ht]
  \begin{center}
    \epsfig{file=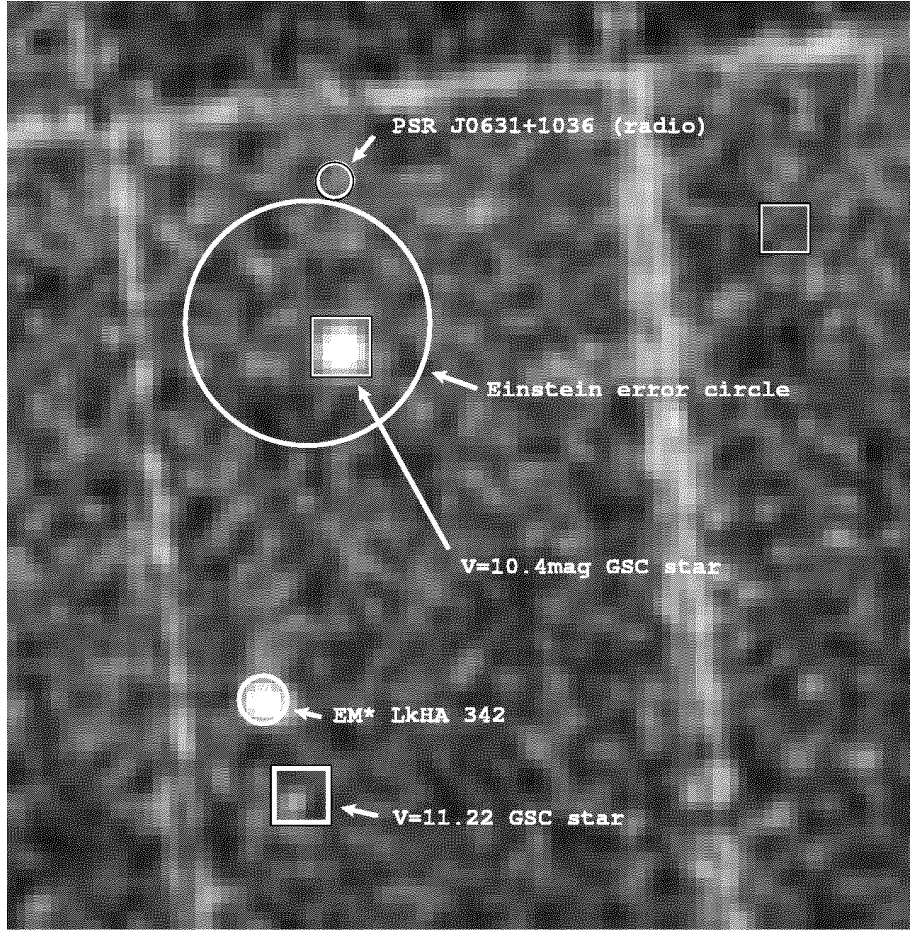, width=12cm}
  \end{center}  
\caption{Annotated EPIC-pn image of the region around PSR J0631+1036. The
  {\it Einstein} error circle of 2E0628.7+1037 clearly contains a bright X-ray
  source. However the radio position of PSR J0631+1036 does not contain any
  detectable X-ray emission. Foreground stars are labeled to show
  correctness of the astrometry}
\label{jkennea-D1:fig1}
\end{figure*}

\begin{figure*}[!ht]
  \begin{center}
    \epsfig{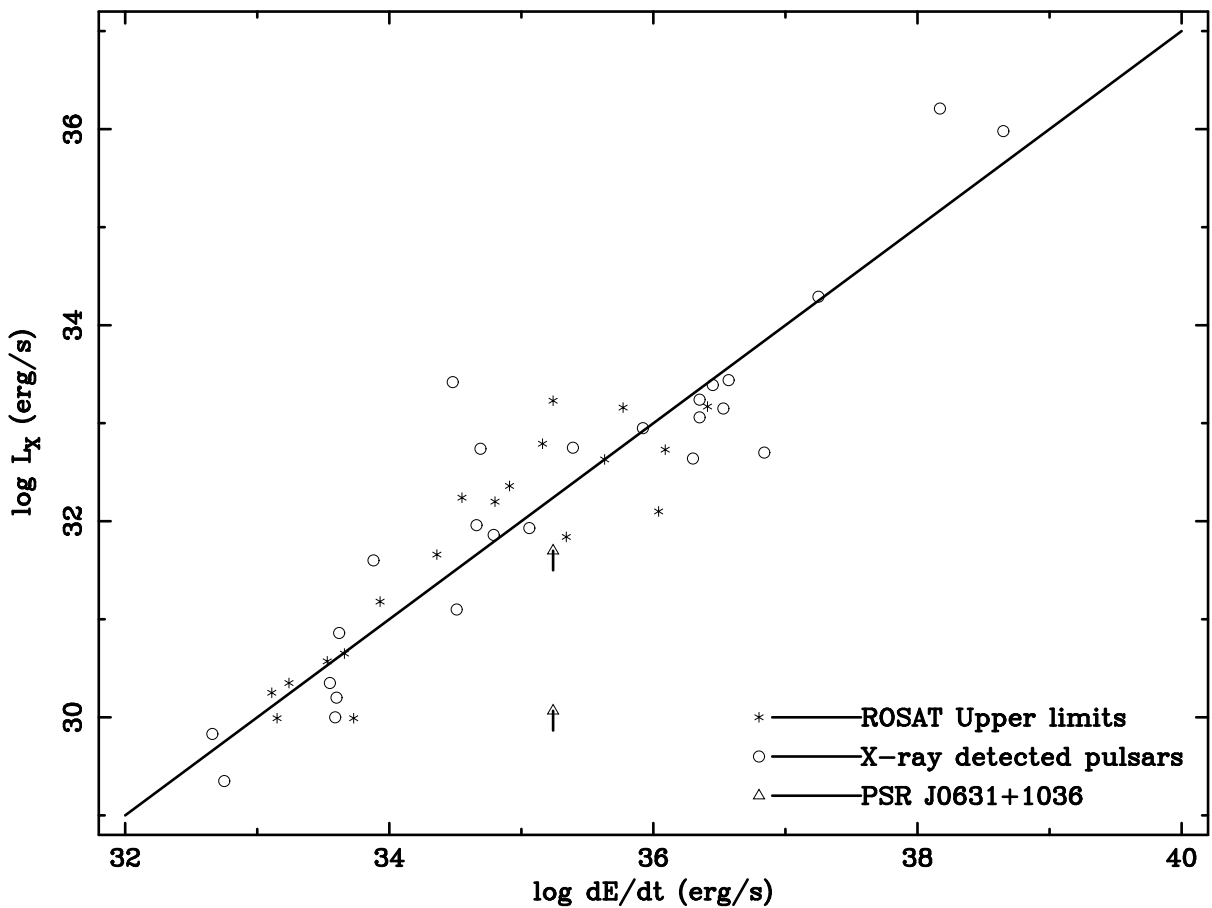}
  \end{center}  
\caption{The Becker and Tr\"umper (1997) empirical relationship between the
  radio measured dE/dt and 0.1-2.4 keV X-ray flux, plotting both X-ray
  detections of pulsars and $3\sigma$ upper limits derived from {\it ROSAT}. The upper
  limits for PSR J0631+1036 from the {\it XMM-Newton} data are marked for
  distances of 6.56kpc (upper arrow) and 1kpc (lower arrow).}
\label{jkennea-D1:fig3}
\end{figure*}

\section{Analysis of the PSR J0631+1036 field}

Figure~\ref{jkennea-D1:fig1} shows the {\it XMM-Newton} EPIC-pn image of
the region around PSR J0631+1036. The image has been annotated with
positions of Simbad catalogue objects. The astrometry of the pn data has
been confirmed by comparing with that of the Optical Monitor data, and has
been found to be consistent with foreground star positions to $\sim1$
arcsecond.  From Figure~\ref{jkennea-D1:fig1} we see that 2E 0628.7+1037 is
detected within the {\it Einstein} error circle, however with {\it
  XMM-Newton}'s improved spatial resolution it is now clear that the radio
position of PSR J0631+1036 is not consistent with the X-ray position of 2E
0628.7+1037, but in fact lies $75''$ north. Analysis of the region around
PSR J0631+1036 shows that there is no X-ray source detected, and the
$3\sigma$ upper limit on X-ray luminosity is: $L_x < 5.0 \times 10^{31}
\mathrm{\ erg/s},$ (0.5--2.0 keV) assuming a distance of 6.56kpc, or $L_x <
1.1 \times 10^{30} \mathrm{\ erg/s},$ (0.5--2.0 keV), assuming a 1 kpc
distance. These upper limits are shown in Figure~\ref{jkennea-D1:fig3}
along with fluxes and {\it ROSAT} upper limits for other pulsars (from
\cite{jkennea-D1:bt97}). For a more detailed discussion of the upper limits
on X-ray emission from PSR J0631+1036 see \cite*{jkennea-D1:m02}.

Comparing this to the previously obtained upper limit from {\it ROSAT} of
$1.7 \times 10^{33}$ erg/s, shows that {\it XMM-Newton} allows us to
improve the upper limit on X-ray emission from PSR J0631+1036 by a factor
of 30 over the upper limit deduced by \cite*{jkennea-D1:bt97}.

Catalogue searches show that the X-ray source previously thought to be
X-ray emission from PSR J0631+1036 is consistent with a catalogued V=10.4
A0 spectral classification. Further analysis of the {\it XMM-Newton}
optical monitor UVW2 data confirms the A0 spectral type of this star. This
star is present in the Tycho catalogue and has a parallax of
53.9+/-26.4mas, corresponding to a distance of 12--36 parsecs. The
estimated luminosity of 2E 0628.7+1037 is consistent with that of X-ray
emission from an A0 star (e.g. \cite{jkennea-D1:p99}), however the errors
on the luminosity are large due to the uncertainty of the distance to the
A0 star.

Spectral fitting of this source with the blackbody plus power-law model
used by \cite*{jkennea-D1:t01} shows that the spectral parameters of 2E
0628.7+1037 are consistent with those derived from the {\it ASCA} data.  However
the {\it XMM-Newton} data shows an improved spectral fit utilising an absorbed
MEKAL model (e.g.  \cite{jkennea-D1:m85}) with a fitted temperature of
kT=0.9+/-0.1 keV and an abundance of $\sim$0.15 of solar.

We have also performed temporal analysis of the area around PSR J0631+1036
and from the detected X-rays from 2E 0628.7+1037. In both cases there is no
significant detection of the 288ms pulsar period.  Simulations show that if
the periodicity was present in 2E 0628.7+1037 at the level reported by
\cite*{jkennea-D1:t01}, it would have been detected with high significance
by {\it XMM-Newton}.

\begin{figure}[!ht]
  \begin{center}
    \epsfig{file=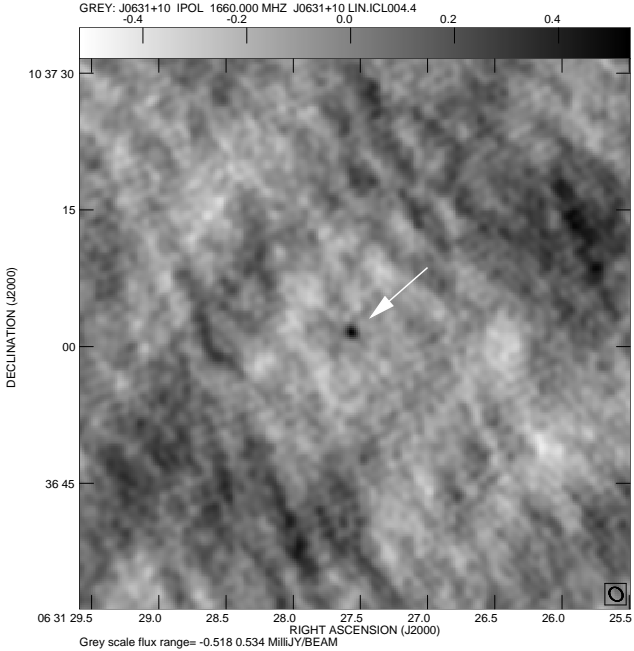, width=8.5cm}
  \end{center}  
\caption{VLA image of PSR J0631+1036, clearly showing a detection of
the pulsar. The coordinates of PSR J0631+1036 derived from these data
are consistent with that of the originally published timing position
in Zepka et al. (1996) and not consistent with any X-ray emission
detected by the EPIC-pn.}
\label{jkennea-D1:fig2}
\end{figure}

\section{Discussion}

Our results show that the association of PSR J0631+1036 and the {\it
  Einstein} source, 2E 0628.7+1037, is not real despite the claimed
detection of the pulsar period in the {\it ASCA} data.

Although the position of the radio and X-ray source are offset by $75''$, a
3 \arcmin extraction radius was used in the {\it ASCA} analysis, and in principle
the signal from the pulsar could be included in the total signal. However
given that the suggested {\it ASCA} modulation had a 45\% pulse fraction, we
would expect that if PSR J0631+1036 were contributing this modulation, it
would be detected by {\it XMM-Newton}.  It is therefore likely that the
period detected by \cite*{jkennea-D1:t01} is a statistical artifact rather
than a real modulation.

Another possibility is that the radio timing position of pulsar is
incorrect. To rule this out we analysed a previously unpublished VLA
observation of PSR J0631+1036. Figure~\ref{jkennea-D1:fig2} shows VLA
images of the pulsar which confirm its position at RA = 06h31m27.5s, Dec =
$+10^o37'01.6"$ (J2000) with a position error $<1"$, which is consistent with
the originally reported radio timing position \cite*{jkennea-D1:z96}.

Proper motion of the pulsar is not a likely explanation of any differences
in position; if the pulsar were at 1kpc this would imply a proper motion of
60,000 km/s (and much higher if at 6.5kpc). Also the X-ray positions of 2E
0628.7+1037 measured in {\it ROSAT} and {\it XMM-Newton} data are consistent within 10"
despite the observations having been taken $\sim 9$ years apart. Therefore
we conclude that the results published by \cite*{jkennea-D1:t01} must be
incorrect.

Given this we need to re-examine the X-ray properties of PSR J0631+1036, as
at present the literature is misleading. If we assume that PSR J0631+1036's
X-ray emission follows the empirical relationship of
\cite*{jkennea-D1:bt97}, $L_x = 10^{-3} {\dot E} \mathrm{\ erg/s}$, then
the expected X-ray luminosity from this source will be: $L_x = 1.7 \times
10^{32} \mathrm{\ erg/s},$ which is close to our upper limit of $5.0 \times
10^{31}$ erg/s (for 6.56kpc).  This upper limit lies $1\sigma$ below the
\cite*{jkennea-D1:bt97} relationship line shown in
Figure~\ref{jkennea-D1:fig3}. We therefore conclude that the non-detection
of the pulsar by {\it XMM-Newton} is consistent with the
\cite*{jkennea-D1:bt97} relationship if the distance to the pulsar is
6.56kpc.

If the distance to pulsar is 1kpc as suggested by \cite*{jkennea-D1:z96},
the X-ray upper limit is now $\sim2$ orders of magnitude lower than
estimated $10^{-3}\dot E$ value, suggesting that the emission is not
following the \cite*{jkennea-D1:bt97} relationship. However without an
accurate measurement of the distance to PSR J0631+1036 this cannot be
considered firm evidence. Therefore further X-ray observations are
encouraged to further constrain the distance and X-ray luminosity of this
pulsar.

\section{Conclusions}

X-rays from PSR J0631+1036 have not been detected by {\it XMM-Newton}.
Therefore the detection of the 288ms pulsar period in {\it ASCA} data by
\cite*{jkennea-D1:t01} is not likely to be correct due to the $75''$
discrepancy between the position of the X-ray source and the radio position
of the pulsar. It is therefore considered very unlikely that the object 2E
0628.7+1037 is related to PSR J0631+1036.

\section{Acknowledgments}

Observations were made with {\it XMM-Newton}, an ESA science mission,
funded by ESA member states and the USA (NASA).  This work was supported by
NASA grant NAG5-7714.

\end{document}